\documentclass[aps,prb,reprint,superscriptaddress,nobibnotes,showkeys,tightenlines]{revtex4-1}
\usepackage{array}  
\usepackage{booktabs} 
\usepackage{comment} 

\usepackage[english]{babel}

\usepackage{amsmath}
\usepackage{graphicx}
\usepackage[colorlinks=true, allcolors=blue]{hyperref}

\begin{document}

\title{Materials Properties Prediction (MAPP):\\ Empowering the prediction of material properties solely based on chemical formulas}
\author{Si-Da Xue}
\affiliation{School for Engineering of Matter, Transport and Energy, Arizona State University, Tempe, AZ 85287, USA}


\author{Qi-Jun Hong}
\email[e-mail:]{ qhong@alumni.caltech.edu}
\affiliation{School for Engineering of Matter, Transport and Energy, Arizona State University, Tempe, AZ 85287, USA}

\begin{abstract}

Predicting material properties has always been a challenging task in materials science. With the emergence of machine learning methodologies, new avenues have opened up. In this study, we build upon our recently developed graph neural network (GNN) approach to construct models that predict four distinct material properties. Our graph model represents materials as element graphs, with chemical formulas serving as the only input. This approach ensures permutation invariance, offering a robust solution to prior limitations. By employing bootstrap methods to train on this individual GNN, we further enhance the reliability and accuracy of our predictions. With multi-task learning, we harness the power of extensive datasets to boost the performance of smaller ones. We introduce the inaugural version of the Materials Properties Prediction (MAPP) framework, empowering the prediction of material properties solely based on chemical formulas. 


\end{abstract}

\maketitle
\section{Introduction}

The accurate prediction of material properties is a demanding and time-consuming endeavor that presents persistent challenges, despite the extensive scientific efforts invested. Traditionally, experimental measurements and computational simulations have played primary roles in this pursuit. However, their reach was constrained by the throughput speed and the scope of chemical systems under investigation. Recent breakthroughs, such as the continuous expansion of material databases \cite{choudhary2020joint,jain2013commentary,kirklin2015open,curtarolo2012aflowlib,hellenbrandt2004inorganic,zakutayev2018open} and the rapid advancement of machine learning algorithms \cite{lecun2015deep,choudhary2022recent,hong2020machine,zhou2020graph}, have revolutionized the landscape of material research. These developments have facilitated the ubiquitous adoption of machine learning models, introducing fresh perspectives to the field of material research and substantially accelerating the process of materials discovery.
As a powerful alternative and complement to physics-based simulation, machine learning offers significant advancements in predicting material properties. Yet, current models in material science face limitations, often requiring specific descriptors and detailed crystal structures for input. The selection of new descriptors can be either a trial and error process or a challenging task demanding an understanding of physical mechanisms. The crystal structure is most likely unknown for an arbitrary chemical formula. Consequently, despite numerous works \cite{jha2018elemnet,zheng2018machine,le2020critical,schmidt2021crystal,allotey2021entropy,stanev_machine_2018,zhang2020robust} that use machine learning for individual material science problems, the broad applicability and adaptability of machine learning remain underdeveloped.

To overcome these limitations, we develop a comprehensive framework that leverages the fundamental principle of using elements as building blocks and chemical composition as the input parameter, which enables the rapid and accurate computation of material properties solely based on chemical formulas. Our framework requires only the chemical formula as an input variable, allowing the prompt calculation of properties associated with any given chemical formula. We harness the capabilities of databases to build models and offer public access, transcending the conventional constraints of limited database size and entry count. 

This approach offers several advantages. First, it eliminates the need for additional input beyond the chemical formula, making it applicable to any chemical formula, which is typically the only a priori known input for new material. Second, since the models can handle any chemical formula, they have the capability to perform queries without being constrained by a limited database with a finite number of data entries. Third, the output value of the model is determined collectively by the entire dataset, rather than a single data point, which potentially minimizes errors. Lastly, the MAPP framework and its models are publicly accessible via the internet, thus empowering individuals without expertise in density functional theory (DFT) or machine learning (ML) to compute material properties rapidly and precisely. This framework has the potential to transform the approaches adopted by modelers for designing products and by experimentalists for utilizing their predictions, thereby revolutionizing materials design and discovery for the future.

\section{Methods}
In the MAPP framework, we achieve the overall goal in three steps. First, we design a unique deep learning architecture that operates on any chemical formula with only the formula itself as the input. With this advantage, material design can rapidly survey the complete chemical space, without requiring any additional information about the material’s properties, as the chemical formula is typically the only \textit{a priori} known input for new material. Second, we build a system to integrate data from various sources \cite{hong2022integrating}, including experiments, DFT, and ML. Using the deep learning model in Step 1 as a foundation, we employ bootstrap aggregation (or cross-validation) and construct ensemble models to not only provide uncertainty quantification but also to detect outliers that can be reviewed and rectified manually. This approach is particularly useful as we assimilate data from various sources such as experiments, DFT, and ML, among others. Third, we build the MAPP framework, characterized by a diverse array of material properties, the potential for iterative enhancement, and the prospect of model integration for systematic improvement. 
Our initial success on the melting temperature \cite{hong2022melting} demonstrates the approach is viable and effective, thus prompting us to further extend our efforts to construct models for predicting four additional material properties, including bulk modulus, volume, heat of fusion, and critical temperature of superconductor. 
Furthermore, the availability of a well-performing model for a given property enables us to surpass the limitation of using only the chemical formula as input. By integrating multiple models and leveraging the well-established model, we can enhance the performance of other models, thus developing an interconnected network of models where a model contributes to the improvement of the others.
\subsection{Data}
We have selected five distinct material properties to showcase the predictive power of our GNN model. These properties span a diverse range of material characteristics, encompassing mechanical, structural, electrical, and thermal properties. Specifically, we focus on melting temperature, heat of fusion, bulk modulus, volume, and superconducting critical temperature. The effective performance of the melting temperature model has been previously established in our prior research \cite{hong2022melting}, which will not be illustrated here. The rest of the four material properties will be discussed. Certain datasets, such as melting temperature, heat of fusion, and superconducting critical temperature, are obtained from experiments. Conversely, datasets for bulk modulus and volume are sourced from DFT computations. The acquisition of datasets of these material properties is facilitated either through standard material science packages \cite{ong_python_2013} or via web-based data crawling techniques.

\subsubsection{Melting temperature and heat of fusion} 
The datasets for both the melting temperature and heat of fusion were collected from a ten-volume compilation of thermodynamic constants of substances, the book ``Thermodynamic properties of individual substances" \cite{glushko1988thermodynamic}, which is openly available in database format \cite{russian_thermal_property_website}. 

The melting temperature dataset contains 9375 materials, among which 982 compounds exhibit high melting temperatures with melting points exceeding 2,000K. While the majority of the dataset is derived from experimental findings, a small portion originates from DFT-based data generated by our first-principle calculation tool, \textit{SLUSCHI} \cite{Hong2012,hong_user_2016}. The dataset underlying this model predominantly consists of compounds with congruent melting temperatures. Hence, the melting temperature generated by this model is interpreted as the higher end of the solidus-liquidus temperature range. This interpretation similarly applies to compounds that decompose prior to melting. It is crucial to understand that this model does not predict solidus or liquidus temperatures.
 
Regarding the heat of fusion, the dataset comprises 774 data points. Upon inspecting the dataset, we identified an anomaly in the material $\mathrm{GdPd_3}$, leading to its manual removal from the dataset.

\subsubsection{Bulk modulus and volume}
The bulk modulus and volume data were queried from the Materials Project database \cite{jain2013commentary}, utilizing the Pymatgen package \cite{ong_python_2013}. To ensure data consistency, we then filtered out less stable structures, applying a threshold of 10 meV energy above the convex hull. Consequently, the bulk modulus dataset contains 4,236 entries, while the volume dataset consists of 49,213 entries. It is noteworthy that, despite materials initially querying spanning all dimensions (0D, 1D, 2D, and 3D materials), our finalized dataset is restricted to materials with 3D crystal structures, given that only these materials provide well-defined volume values. We note that the bulk modulus and volume data are based on DFT calculations at absolute zero.
\subsubsection{Superconducting critical temperature}
In order to develop a universal model for predicting the superconducting critical temperature $T_c$ with a wide range of $T_c$ value, we utilized a comprehensive dataset from a previous study \cite{hamidieh_data-driven_2018}. This dataset encompasses a wide range of superconductors, including cuprate-based, iron-based, and Bardeen-Cooper-Schrieffer (BCS) theory superconductors, which originate from the Supercon material Database, maintained by the National Institute for Materials Science (NIMS) in Japan \cite{Supercon_dataset}. This database is the most extensive and widely employed resource for data-driven research on superconductors, and it has been extensively utilized in previous studies 
 \cite{stanev_machine_2018,hamidieh_data-driven_2018,zeng_atom_2019,konno_deep_2021}.



\subsection{Model architecture}
\subsubsection{Element embedding}
\begin{figure*}
\centering
\includegraphics[width=1\textwidth]{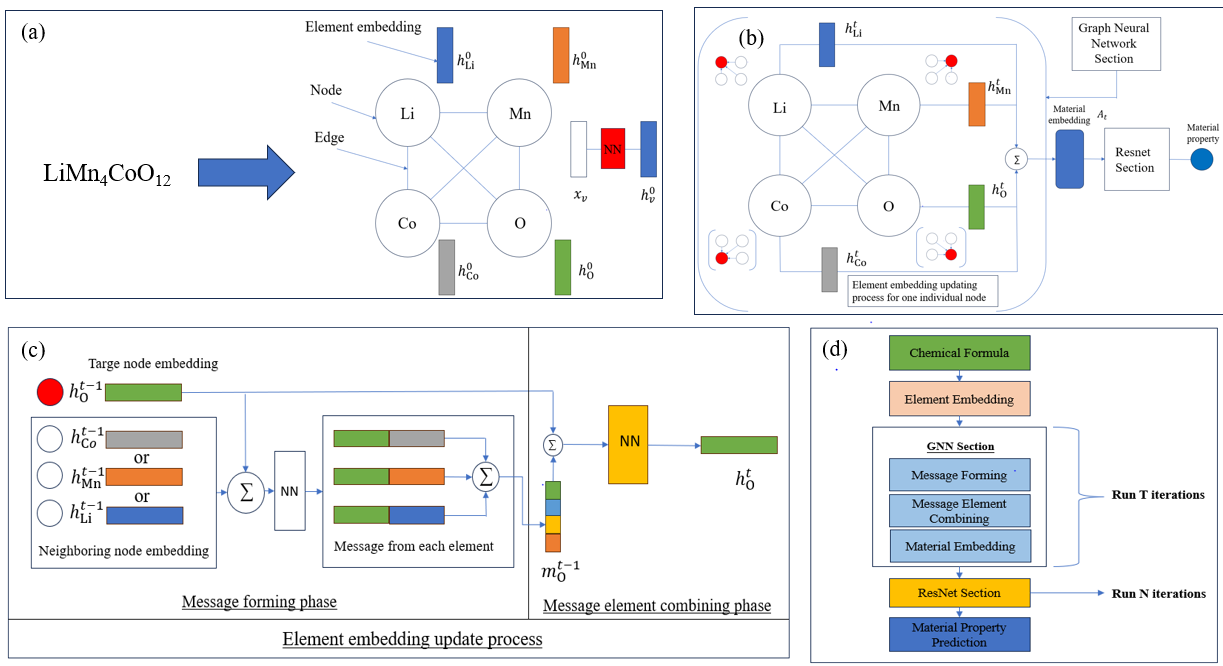}
\caption{\label{fig:model_architecture_fig_paper}(a) Converting a chemical formula to an element graph, (b) The overall illustration of the graph neural network section and ResNet section, (c) The detailed process of the element embedding update process, including the message forming phase and the message element combining phase, (d) The overall workflow of an end-to-end graph neural network deep learning framework. The illustration shows a direct mapping from chemical formulas to material properties.}
\end{figure*}

In our approach, the material's chemical formula is depicted as a fully connected element graph $G_e$. Each element graph $G_e$ comprises nodes and edges, defined as $V$ and $E$ respectively. Figure \ref{fig:model_architecture_fig_paper} (a) shows the process of converting the chemical formula into an element graph. To exemplify, the material $\mathrm{Li_7Mn_4CoO_{12}}$ is visualized as an element graph with nodes corresponding to elements Lithium, Manganese, Cobalt, and Oxygen. Each node represents a specific element, connected to each neighboring node through a single edge. Each edge symbolizes the path for information exchange between paired elements. Specifically, the edges have no features and are treated equally, only indicating the connection of the information-exchanging route. In constructing the element graph, each element within the chemical formula is associated with a node feature vector. Each node feature vector has 14-dimensional features, composed of elemental properties, such as atomic mass, atomic number, melting temperature, boiling temperature, electronegativity, \textit{etc,} as well as the composition of each element. The composition serves as an indicator of the relative importance of each element within the chemical formula of the material.

The node feature vector is denoted as $x_v$, with $v$ symbolizing a specific node within the node-set, $V$. To ensure the permutation invariance for all elements in the subsequent graph neural network section, each $x_v$ undergoes nonlinear transformation through an identical fully connected neural network, utilizing the same activation function. This transformation is expressed as:  $h_v^{0} = \mathrm{ReLU}(W^{0}x_v + b^{0}), v \in V$. In this equation, $\mathrm{ReLU}$ denotes the rectified linear unit activation function. The term $h_v^{0}$ signifies the initial element embedding prior to the graph neural network layer, while the superscript indicates the specific layer within the graph neural network. Moreover, $W^{0}$ and $b^{0}$ denote the weight and bias parameters of the neural network. 
\subsubsection{Graph neural network section}
After creating the element graph based on chemical formula, the graph neural network model is used to transform element embeddings and subsequently, the embedding of the whole element graph in order to capture more physical insights from the material and to perform the material property prediction better. Our task is to predict the material property using the element graph, as shown in Figure \ref{fig:model_architecture_fig_paper} (b), framing this as a graph-level regression task. Within the graph-level prediction task, the objective of the GNN is to learn a representation of the entire graph, the so-called material embedding $A$ in Figure \ref{fig:model_architecture_fig_paper} (b). This is achieved by iteratively updating the node embeddings based on the neighboring information, also termed as the message in literature \cite{gilmer2017neural} and then aggregating the individual node representations to form the material representation, which is used to perform the regression. In the following paragraphs, the element embedding update process and the material embedding update process will be elaborated. 

As shown in Figure \ref{fig:model_architecture_fig_paper} (b), element embedding of the target node is updated by incorporating messages from all the neighboring elements. This inference bias arises from the understanding that the information from the neighboring nodes is not only relevant to the target node but also enriches the target node's information. 

There are generally two phases in the element embedding update process, the message forming phase and the message element combining phase, as is shown in Figure \ref{fig:model_architecture_fig_paper} (c). The message is created based on both the embedding of the neighboring node and the current node and has the general form shown below:
\begin{equation}
m_v^{t-1} = \sum_{u \in N(v)}^{|V|-1} M_t\left(h_v^{t-1}, h_u^{t-1}\right),
\end{equation}
where $m_v^{t-1}$ is the message for target node $v$ at iteration $t-1$, $N(v)$ is the neighboring node set of $v$, $|V|$ is the number of the nodes in the entire graph. $h_v^{t-1}$ and $h_u^{t-1}$ denote the element embeddings of target node $v$ and neighboring node $u$ at iteration $t-1$. The message function $M_t$ can take multiple forms, for example, adding, averaging, concatenating, \textit{etc}, which all serve as a way to preserve the information of the neighboring node. The target node's overall message is updated by aggregating messages from all of the neighboring nodes, which preserves the properties of permutation invariance of the chemical formula. 
In this study, the specific form of message function is

\begin{equation}
m_v^{t-1} = \sum_{u \in N(v)}^{|V|-1} \Bigg[ \mathrm{ReLU}\Bigg( W^{t-1}\mathrm{Add}\left(h_v^{t-1},h_u^{t-1}\right)+b^{t-1} \Bigg) \Bigg],
\end{equation}
where $W^{t-1}$ and $b^{t-1}$ are the weight and bias of the neural network at the $t-1$ iteration. Figure \ref{fig:model_architecture_fig_paper} (c) illustrates the message-forming procedures. The messages from elements Co, Mn, and Li are first created by adding the element embeddings of the target node $h_\mathrm{O}$ and neighboring node $h_u$. Then all the individual messages are passed through an identical neural network layer. An overall message $m_\mathrm{O}$ is formed by aggregating all the individual messages.

After generating the message $m_v^{t-1}$, $h_v$ is updated by summing the message vector $m_v^{t-1}$ with the node embedding from previous layer $h_v^{t-1}$,
\begin{equation}
h_v^{t} = \mathrm{ReLU}\left(W^{t-1}\mathrm{Add}\left(h_v^{t-1}, m_v^{t-1}\right)+b^{t-1}\right).
\end{equation}

The material embedding $A^{t}$ for layer $t$ is constructed by aggregating all element embeddings, a process illustrated in Figure \ref{fig:model_architecture_fig_paper} (b),
\begin{equation}
A^{t} = \sum_{v \in V}^{|V|} h_v^{t}.
\end{equation}

The element and material embedding update process will last for $T$ iterations, which is a hyperparameter to tune.

In our architecture, we record the material embeddings $A$ from every iteration, accumulating them in part to construct the final material embedding, $A^\mathrm{final}$. In general, $A^\mathrm{final}$ may be constituted in two ways. It could either be the material embedding $A^{T}$ derived from the final iteration within the graph neural network, or an aggregation of the material embedding $A$ created during the $T$-th iteration along with those from earlier iterations. This method helps in conserving information from preceding GNN layers, potentially mitigating the prevalent issue of over-smoothing often encountered in GNN applications.


\subsubsection{Property prediction section with ResNet architecture}
After generating a material embedding through the GNN section, the material embedding is fed into the fully connected layers with ResNet architecture \cite{he_deep_2016} for further nonlinear transformation. The ResNet layer has the advantage over the plain dense layer in that it can mitigate the exploding and vanishing gradient issue, therefore, guaranteeing that the model can converge to the optimal solution easier. After $N$ number of ResNet transformations and a regression neural network layer, a final value of the predicted material property is given by the model. 

To summarize the aforementioned model architecture and material prediction process, our GNN model's standard workflow for predicting material properties is depicted in Figure \ref{fig:model_architecture_fig_paper} (d), demonstrating the comprehensive, end-to-end capability of the GNN. In this model, the chemical formula, as the sole input, is encoded as a series of elemental embeddings. Within the GNN layers, these elemental embeddings undergo $T$ iterations of updates, aggregating to form an ultimate material embedding. This embedding is subsequently employed for regression within the ResNet block. 
\subsubsection{Ensemble model and uncertainty estimation}


To further increase the model accuracy, quantify prediction uncertainty, detect outliers, and perform comprehensive data analysis, we propose an ensemble model based on the bootstrap method. The model consists of 30 independent GNN models trained on different sampling from the original dataset, with a final model performance derived from aggregating the performance metrics of 30 individual models, such as coefficient of determination ($R^2$ score), root-mean-square error (RMSE), and mean absolute error (MAE). The training data for each individual model is randomly sampled using the bootstrap method from the original dataset. The testing set is composed of data not included in each sampling process. 

The ensemble model can increase the robustness of the individual deep learning model. The diversity of the model can be ensured because all the models are trained on a different subset of the original dataset. This diversity can potentially improve the model's accuracy since the errors made by different models may be canceled from each other when aggregated. Moreover, the overfitting tends to be averaged out as each individual model in the ensemble is trained on a different subset of the original dataset, which may help the model better generalize to the unseen data. 

Additionally, the utilization of an ensemble model allows for the quantification of uncertainty and facilitates a comprehensive analysis of outliers within the dataset. By ensuring the inclusion of every data point in both training and testing sets at least once, we establish a systematic basis for thoroughly assessing the model's predictive performance for each individual material. This approach guarantees that the training set closely represents the overall distribution of the original dataset, thereby providing a robust evaluation of the model's performance under various scenarios. 
By analyzing the statistics of the predicted errors, specifically the mean, median, and maximum differences between predicted and actual values, we can identify materials that exhibit significant deviations in both the training and testing phases. Such anomalies in the data may necessitate a closer examination to determine the causes of such anomalies. These inconsistencies could stem from incorrect data entries, requiring their removal from the original dataset to enhance accuracy. Alternatively, they might represent unique statistical distributions that our current model fails to recognize, indicating a need for more sophisticated data handling or model adjustment to accommodate these exceptions.

\section{Results and discussion}
\subsection{Bulk modulus}
The bulk modulus, denoted as $k_{\text{vrh}}$ quantifies a material's resistance to uniform compression. Understanding the bulk modulus is essential for researchers investigating the elastic properties of materials. Developing a deep learning model to predict the bulk modulus could facilitate the discovery of ultra-compressible materials. \cite{mansouri_tehrani_machine_2018,kaner_designing_2005}. 
\begin{figure*}
\centering
\includegraphics[width=1\textwidth]{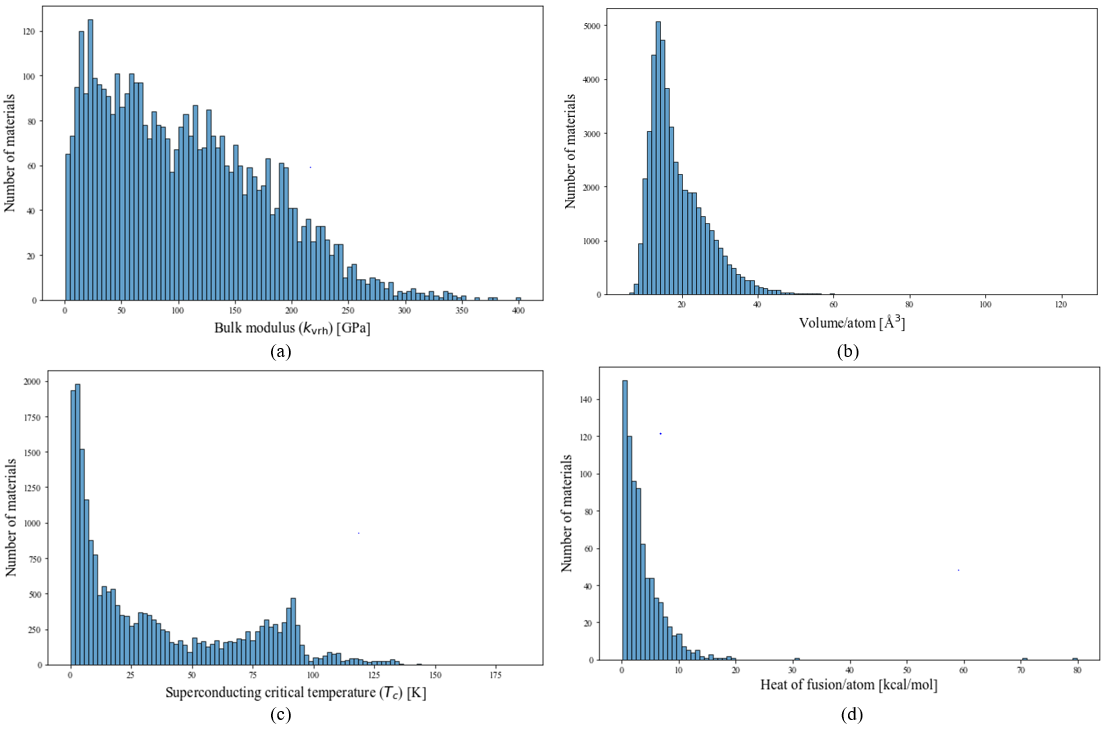}
\caption{\label{fig:data_distribution_unique}The histograms of (a) bulk modulus ($k_{\mathrm{vrh}}$) [GPa], (b) volume/atom [$\text{\r{A}}^3$], (c) Superconducting critical temperature ($T_c$) [K], (d) heat of fusion [kcal/mol].}
\end{figure*}

The data distribution of the bulk modulus dataset used in our work is shown in Figure \ref{fig:data_distribution_unique} (a). 
Using the aforementioned element feature generation method and modeling the material as an element graph, a bulk modulus model was trained with 4236 bulk modulus data entries derived from DFT calculations. In this study, various hyperparameters were utilized, including the number of neurons in a hidden layer, the number of ResNet layers, the number of graph neural network layers, dropout rate, batch size, \textit{etc.} 

\begin{figure*}
\centering
\includegraphics[width=1\textwidth]{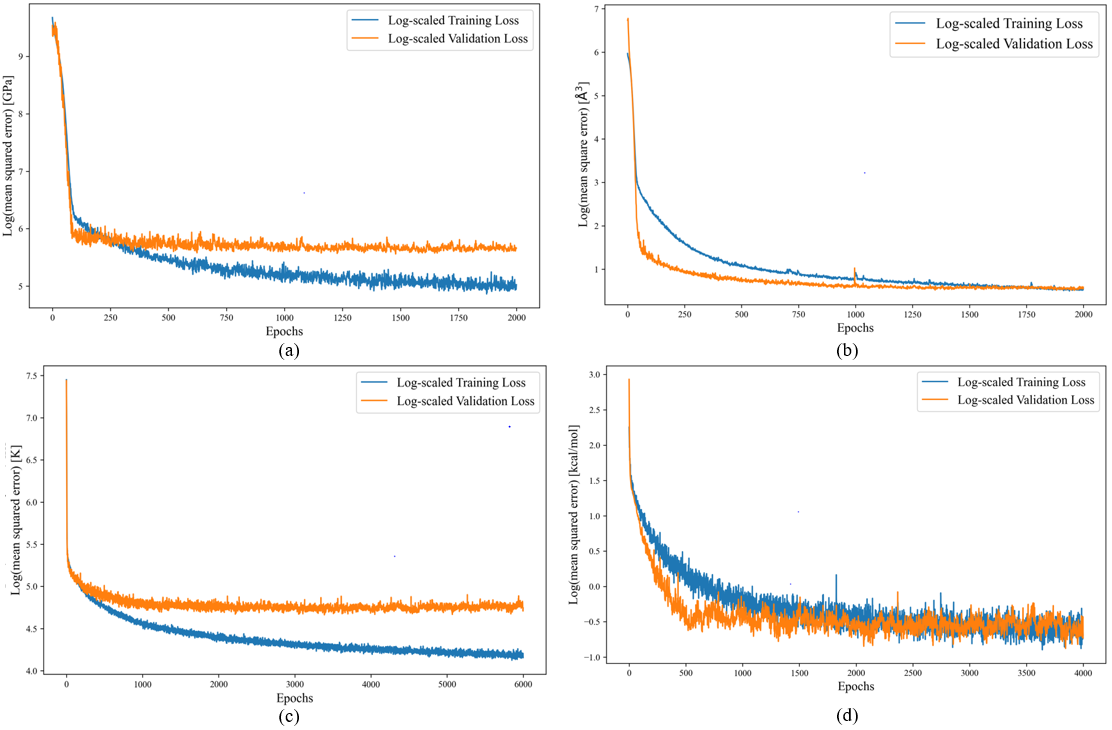}
\caption{\label{fig:Loss_curve_all_fontsize14_unique}The log-scaled training and validation loss functions of (a) bulk modulus ($k_{\mathrm{vrh}}$) [GPa], (b) volume/atom [$\text{\r{A}}^3$], (c) superconducting critical temperature ($T_c$) [K], (d) heat of fusion [kcal/mol]. The loss functions used in this work are mean squared error.}
\end{figure*}

\begin{figure*}
\centering
\includegraphics[width=0.7\textwidth]{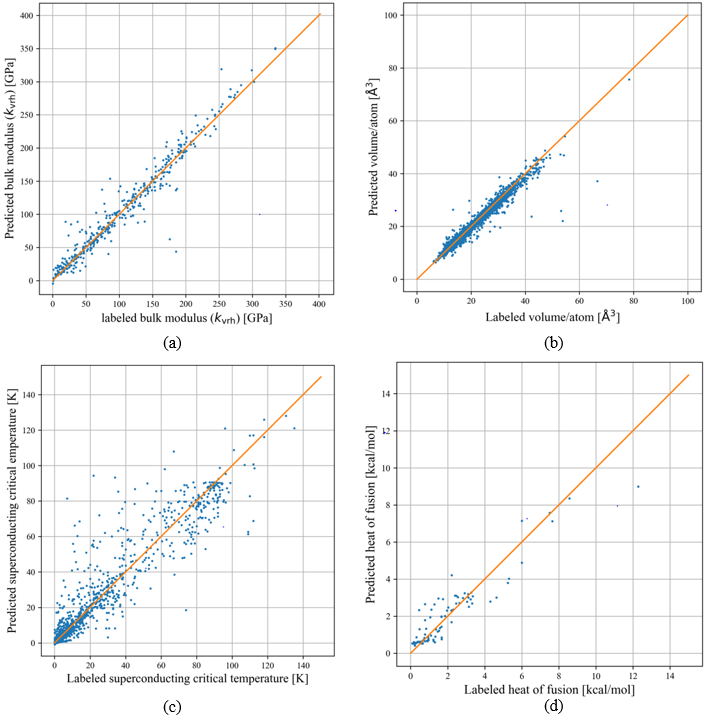}
\caption{\label{fig:parity_plot_all}The parity plots of (a) bulk modulus ($k_{\text{vrh}}$) [GPa], (b) volume/atom [$\text{\r{A}}^3$], (c) superconducting critical temperature ($T_c$) [K], (d) heat of fusion/atom [kcal/mol]. The labeled values and the predicted values for specific material properties are shown in the parity plots.}
\end{figure*}

According to Figure \ref{fig:Loss_curve_all_fontsize14_unique} (a), after 2000 epochs, the model's loss function converges. The coefficient of determination ($R^2$ score) for a single GNN model is 0.95 for the testing set. The parity plot for the testing bulk modulus dataset is shown in Figure \ref{fig:parity_plot_all} (a), which showcases the difference between the labeled bulk modulus values and the predicted bulk modulus values. The RMSE and MAE for the test set are 17.04 and 9.96 GPa, respectively. Moreover, we trained an ensemble model of 30 individual models with the same hyperparameters but with different training and testing sets generated from the bootstrap method. The model achieved a testing $R^2$, RMSE and MAE of 0.93, 19.41 GPa, and 11.2 GPa respectively. A summary of the performance of the single and ensemble bulk modulus model, $R^2$ score, RMSE, and MAE is shown in Table ~\ref{tab:widgets}. 

\begin{table*}
\caption{\label{tab:widgets}Performances for the superconducting critical temperature, volume, bulk modulus, and heat of fusion models. Both the single model and the ensemble model are included.}
\renewcommand{\arraystretch}{1.6} 
\setlength{\tabcolsep}{5pt} 

\begin{tabular*}{\textwidth}{@{\extracolsep{\fill}}|c|ccc|ccc|}
\hline
Model type & \multicolumn{3}{c|}{Single model} & \multicolumn{3}{c|}{Ensemble model} \\ 
\hline
Material properties & \multicolumn{1}{c|}{$R^2$ score} & \multicolumn{1}{c|}{RMSE} & MAE & \multicolumn{1}{c|}{$R^2$ score} & \multicolumn{1}{c|}{RMSE} & MAE \\ 
\hline
Bulk modulus ($k_{\mathrm{vrh}}$) [GPa] & 0.95 & 17.04 & 9.96 & 0.93 & 19.41 & 11.2 \\ 
\hline
Unit cell volume/atom [$\text{\r{A}}^3$] & 0.97 & 1.56 & 0.65 & 0.97 & 1.36 & 0.84 \\ 
\hline
Superconducting critical temperature ($T_c$) [K] & 0.91 & 10.16 & 6.91 & 0.88 & 12.64 & 7.54 \\ 
\hline
Heat of fusion/atom [kcal/mol] & - & - & - & 0.70 & 1.15 & 0.74\\ 
Heat of fusion/atom (multi-task learning) [kcal/mol] & - & - & - & 0.74 & 1.01 &  0.67  \\ 
\hline
\end{tabular*}
\end{table*}



\subsection{Volume}

The volume data displays significant variability, with a minimum value of 11 and a maximum of 7132 $\text{\r{A}}^3$. 
This diversity poses difficulties in model fitting. To address this challenge, we employ volume per atom as the target label, leading to a more constrained predictive range. After calculating the volume per atom value, the data distribution is shown in Figure \ref{fig:data_distribution_unique} (b). After training for 2000 epochs, the loss function of the single GNN model converges to an optimal level, as shown in Figure \ref{fig:Loss_curve_all_fontsize14_unique} (b), achieving a $R^2$ score of 0.97 for the testing set. The RMSE and MAE of the single model are 1.56 and 0.65 $\text{\r{A}}^3$. The ensemble model achieves an average $R^2$ score, RMSE, and MAE of 0.97, 1.36 $\text{\r{A}}^3$ and 0.84 $\text{\AA}^3$ for the testing set across 30 distinct models. A more detailed result of the model's performance can be seen in the table ~\ref{tab:widgets}.

\subsection{Superconducting critical temperature}

The critical temperature $T_c$ dataset can be divided into three groups: 2,339 iron-based, 10,838 copper-based, and 8,535 other types of superconductors. The category labeled as ``other superconductors" primarily consists of materials explained by the BCS theory, which is effective for low-temperature superconductors. However, this theory does not adequately explain the superconducting behavior in high-temperature superconductors like copper-based and iron-based ones, as highlighted in previous studies \cite{bardeen_theory_1957,mann_high-temperature_2011}. In the pursuit of high-temperature superconductors, deep learning models play a crucial role. These models aid in expanding the chemical space for analysis through high-throughput computational screening, thus speeding up the discovery of new materials. Once materials with a high potential for high $T_c$ are identified, experimental validation can be conducted. The $T_c$ model effectively directs the material synthesis efforts on a more promising subset of high-temperature superconductors.
The $T_c$ data distribution is shown in Figure \ref{fig:data_distribution_unique} (c). The threshold for the high $T_c$ is 30 K \cite{bednorz1986possible}. By inspecting the $T_c$ dataset, 8848 superconductors have $T_c$ values above this threshold. 
The model achieved convergence after approximately 6000 epochs, as depicted in Figure \ref{fig:Loss_curve_all_fontsize14_unique} (c).
Following hyperparameter tuning, the best single $T_c$ model exhibited a $R^2$ score of 0.91 for the testing set. Additionally, the RMSE and MAE are found to be 10.16 and 6.91 K for the testing set. The parity plot of the $T_c$ model is shown in Figure \ref{fig:parity_plot_all} (c).

After creating a single GNN model, we trained an ensemble model based on 30 distinct models using data generated via the bootstrap method. The average model performance metrics are listed below: the testing $R^2$ score is 0.88, while the RMSE and the MAE are 12.64 and 7.54 K, respectively. The single model and ensemble model performance metrics are shown in Table ~\ref{tab:widgets}.

\subsection{Heat of fusion}
The heat of fusion dataset currently contains a significantly smaller 742 data points. Within this dataset, the label employed is the heat of fusion per number of atoms in the chemical formula, rather than the absolute heat of fusion values. This normalization shifts the model's focus towards discerning the average contribution of each atom to the heat of fusion, mitigating the influence of the compound's size.  The distribution of this data is shown in Figure \ref{fig:data_distribution_unique} (d). The $R^2$ score, RMSE, and MAE of the heat of fusion model are 0.70, 1.15 kcal/mol, and 0.74 kcal/mol, respectively, for the testing set. 

Enhancing the accuracy of this model can be achieved by either collecting more data \cite{Hong2012, hong_user_2016} or harnessing the multi-task learning technique \cite{crawshaw2020multi} to facilitate the training of the target material property, in this case, the heat of fusion. The multi-task learning leverages an auxiliary material property characterized by a larger dataset size, better data quality, and an inherent correlation with the target property, to aid the training process of the target task.

\begin{figure}
\centering
\includegraphics[width=0.49\textwidth]{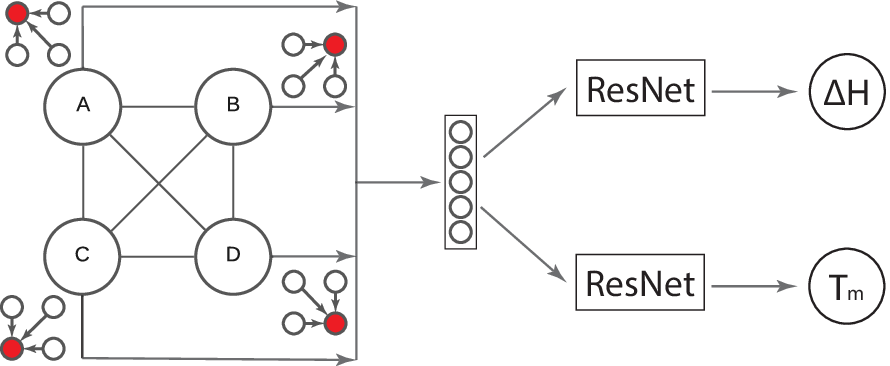}
\caption{\label{fig:multitask}Diagram of multi-task learning utilizing the large melting temperature dataset to augment the training of the significantly smaller heat of fusion dataset.}
\end{figure}

We utilized the melting temperature dataset, discussed previously \cite{hong2022melting}, as an auxiliary task to facilitate the training of the heat of fusion task. This study adopted the hard parameter-sharing multi-task learning architecture, as illustrated in Fig. \ref{fig:multitask}.
This architecture employs shared GNN layers and weights across both material properties, effectively expanding the material representation space through simultaneous training on both the melting temperature and heat of fusion datasets. Following these shared layers, the model utilizes two separate ResNet architectures tailored to the regression tasks corresponding to each property. 

The loss curve for the heat of fusion model after applying the multi-task learning is shown in Figure \ref{fig:Loss_curve_all_fontsize14_unique} (d). the model starts to converge at the 4000 epochs. After employing the multi-task learning methodologies, the testing $R^2$ score for the heat of fusion model improved from 0.70 to 0.74. The parity plot of the heat of fusion model after employing multi-task learning is shown in Figure \ref{fig:parity_plot_all} (d). The testing RMSE and MAE are 1.01 and 0.67 kcal/mol.
This progress in the performance highlights the ability of multi-task learning to enhance the training of a machine learning model. Furthermore, it illustrates an alternative way for expanding the model's representational capacity without necessarily increasing the number of data points.

Due to the small size of the dataset, only the performance metrics of the ensemble model are presented, as the performance of a single model exhibits large fluctuation.

\subsection{Discussion} 
Our MAPP framework has been applied to five material properties, four of which are illustrated in this work while the melting temperature is presented thoroughly in our previous articles \cite{hong2022melting,HONG2022_CMS,HONG2022_CALPHAD}. All the above-mentioned models have been deployed and are publicly accessible via our group's website. Notably, our model is distinguished by its input simplicity, requiring only the chemical formula, and by its robust model performance, merit deriving largely from the graph neural network's inherent local information consolidation capability. 

Our ensemble model is capable of uncertainty quantification, which serves to identify materials that yield high prediction errors, potentially flagging outliers. We employ the bootstrap method to segregate data into training and testing subsets, also called in-bag and out-of-bag sets. This process creates 30 independent models, and the final performance of the ensemble model is determined by the aggregated outputs of these individual models. The employment of 30 separate bootstrap iterations ensures comprehensive coverage, with each data point in the original dataset featuring in the testing set at least once. Consequently, the in-bag and out-of-bag errors serve as metrics for assessing our model's accuracy in predicting the properties of specific materials. 

We showcase the potential of multi-task learning in our study by utilizing the sizable melting temperature dataset to augment the training of the significantly smaller heat of fusion dataset, especially beneficial in scenarios of limited data availability. This approach has resulted in considerable enhancements in model accuracy. Despite these advancements, the model's performance has not yet reached an ideal level, which we attribute primarily to the data's limited quantity and suboptimal quality, rather than to any deficiencies in the model's design.

Alternatively, active learning could be employed to expand the dataset. This method would involve leveraging a first-principles calculation pipeline such as the SLUSCHI package \cite{hong_user_2016}, tailored for computing high-temperature material properties using DFT. Through this pipeline, we generate a more comprehensive array of data points for properties such as melting temperature and heat of fusion. Such enrichment of the dataset holds the promise of significantly enhancing the deep learning model's accuracy. Nevertheless, the implementation of this approach falls outside the scope of the current study.

Nonetheless, it is important to acknowledge the limitations of our existing model framework. Our current model relies solely on the chemical formula as inputs. While it simplifies the input process and maximizes general applicability, it also brings certain limitations: it is unable to differentiate between polymorphs — materials sharing the same formula but exhibiting distinct crystal structures. For example,  it cannot distinguish between diamond and graphite. To overcome this limitation, we will consider incorporating crystal structure information into the inputs, which we plan to undertake once we have developed a robust model capable of crystal structure prediction.
\section{Conclusion}

We introduce the MAPP framework, an extensive platform for material property prediction, capable of delivering comprehensive material data based solely on chemical formula input. Utilizing our generic graph neural network approach, we have successfully developed five robust models for predicting solid-state material properties. These models, solely based on the chemical formulae, bypass the need for manual feature engineering. With minimal input of physical information, our models enable us to explore the entire high-dimensional chemical space, exhibiting good performance and remarkable adaptability across various material property prediction tasks, and demonstrating their potential in combinatorial material screenings. We further enhanced the platform by incorporating ensemble models, which allow for the systematic identification of uncertainties and outliers, improving overall performance. 
This paper also sheds light on our strategy of harnessing inter-property correlation to enrich individual model learning. We demonstrate how the multi-task learning approach substantially improves the model's performance in predicting heat of fusion. 

We have designed a user-friendly web application \cite{MAPP} for rapid predictions of material properties. This platform allows public access to our models, enabling users to obtain both material property predictions and associated uncertainties. In addition, we build Application Programming Interfaces (APIs) \cite{MAPP_api} equipped with batch processing capabilities. Through these tools, users can:
\begin{enumerate} 
\item Evaluate materials properties across large datasets,
\item Run interactive simulations for the design and discovery of materials with extreme properties, and 
\item Include materials properties as new features for their models.
\end{enumerate}
Based on traffic analysis, our websites and APIs have so far performed over 7,000 and 300,000 calculations for our users, respectively. The melting temperature model is featured by the Materials Project team at their webpage \cite{jain2013commentary}.

\section{Acknowledgements}

This research was funded by the Army Research Office (ARO) of the U.S. Department of Defense under the Multidisciplinary University Research Initiative W911NF-23-2-0145. We also appreciate the start-up funding provided by the School of Engineering for Matter, Transport, and Energy (SEMTE) and the use of Research Computing at Arizona State University.

\bibliographystyle{unsrt}
\bibliography{reference_for_paper_MAPP}

\begin{thebibliography}{10}

\bibitem{choudhary2020joint}
Kamal Choudhary, Kevin~F Garrity, Andrew~CE Reid, Brian DeCost, Adam~J Biacchi,
  Angela~R Hight~Walker, Zachary Trautt, Jason Hattrick-Simpers, A~Gilad Kusne,
  Andrea Centrone, et~al.
\newblock The joint automated repository for various integrated simulations
  (jarvis) for data-driven materials design.
\newblock {\em npj Computational Materials}, 6(1):173, 2020.

\bibitem{jain2013commentary}
Anubhav Jain, Shyue~Ping Ong, Geoffroy Hautier, Wei Chen, William~Davidson
  Richards, Stephen Dacek, Shreyas Cholia, Dan Gunter, David Skinner, Gerbrand
  Ceder, et~al.
\newblock Commentary: The materials project: A materials genome approach to
  accelerating materials innovation.
\newblock {\em APL Materials}, 1(1), 2013.

\bibitem{kirklin2015open}
Scott Kirklin, James~E Saal, Bryce Meredig, Alex Thompson, Jeff~W Doak,
  Muratahan Aykol, Stephan R{\"u}hl, and Chris Wolverton.
\newblock The open quantum materials database (oqmd): assessing the accuracy of
  dft formation energies.
\newblock {\em npj Computational Materials}, 1(1):1--15, 2015.

\bibitem{curtarolo2012aflowlib}
Stefano Curtarolo, Wahyu Setyawan, Shidong Wang, Junkai Xue, Kesong Yang,
  Richard~H Taylor, Lance~J Nelson, Gus~LW Hart, Stefano Sanvito, Marco
  Buongiorno-Nardelli, et~al.
\newblock Aflowlib. org: A distributed materials properties repository from
  high-throughput ab initio calculations.
\newblock {\em Computational Materials Science}, 58:227--235, 2012.

\bibitem{hellenbrandt2004inorganic}
Mariette Hellenbrandt.
\newblock The inorganic crystal structure database (icsd)—present and future.
\newblock {\em Crystallography Reviews}, 10(1):17--22, 2004.

\bibitem{zakutayev2018open}
Andriy Zakutayev, Nick Wunder, Marcus Schwarting, John~D Perkins, Robert White,
  Kristin Munch, William Tumas, and Caleb Phillips.
\newblock An open experimental database for exploring inorganic materials.
\newblock {\em Scientific Data}, 5(1):1--12, 2018.

\bibitem{lecun2015deep}
Yann LeCun, Yoshua Bengio, and Geoffrey Hinton.
\newblock Deep learning.
\newblock {\em Nature}, 521(7553):436--444, 2015.

\bibitem{choudhary2022recent}
Kamal Choudhary, Brian DeCost, Chi Chen, Anubhav Jain, Francesca Tavazza, Ryan
  Cohn, Cheol~Woo Park, Alok Choudhary, Ankit Agrawal, Simon~JL Billinge,
  et~al.
\newblock Recent advances and applications of deep learning methods in
  materials science.
\newblock {\em npj Computational Materials}, 8(1):59, 2022.

\bibitem{hong2020machine}
Yang Hong, Bo~Hou, Hengle Jiang, and Jingchao Zhang.
\newblock Machine learning and artificial neural network accelerated
  computational discoveries in materials science.
\newblock {\em Wiley Interdisciplinary Reviews: Computational Molecular
  Science}, 10(3):e1450, 2020.

\bibitem{zhou2020graph}
Jie Zhou, Ganqu Cui, Shengding Hu, Zhengyan Zhang, Cheng Yang, Zhiyuan Liu,
  Lifeng Wang, Changcheng Li, and Maosong Sun.
\newblock Graph neural networks: A review of methods and applications.
\newblock {\em AI open}, 1:57--81, 2020.

\bibitem{jha2018elemnet}
Dipendra Jha, Logan Ward, Arindam Paul, Wei-keng Liao, Alok Choudhary, Chris
  Wolverton, and Ankit Agrawal.
\newblock Elemnet: Deep learning the chemistry of materials from only elemental
  composition.
\newblock {\em Scientific Reports}, 8(1):17593, 2018.

\bibitem{zheng2018machine}
Xiaolong Zheng, Peng Zheng, and Rui-Zhi Zhang.
\newblock Machine learning material properties from the periodic table using
  convolutional neural networks.
\newblock {\em Chemical science}, 9(44):8426--8432, 2018.

\bibitem{le2020critical}
Thanh~Dung Le, Rita Noumeir, Huu~Luong Quach, Ji~Hyung Kim, Jung~Ho Kim, and
  Ho~Min Kim.
\newblock Critical temperature prediction for a superconductor: A variational
  bayesian neural network approach.
\newblock {\em IEEE Transactions on Applied Superconductivity}, 30(4):1--5,
  2020.

\bibitem{schmidt2021crystal}
Jonathan Schmidt, Love Pettersson, Claudio Verdozzi, Silvana Botti, and
  Miguel~AL Marques.
\newblock Crystal graph attention networks for the prediction of stable
  materials.
\newblock {\em Science Advances}, 7(49):eabi7948, 2021.

\bibitem{allotey2021entropy}
Johannes Allotey, Keith~T Butler, and Jeyan Thiyagalingam.
\newblock Entropy-based active learning of graph neural network surrogate
  models for materials properties.
\newblock {\em The Journal of Chemical Physics}, 155(17), 2021.

\bibitem{stanev_machine_2018}
Valentin Stanev, Corey Oses, A.~Gilad Kusne, Efrain Rodriguez, Johnpierre
  Paglione, Stefano Curtarolo, and Ichiro Takeuchi.
\newblock Machine learning modeling of superconducting critical temperature.
\newblock {\em npj Computational Materials}, 4(1):29, 2018.

\bibitem{zhang2020robust}
Jiaxin Zhang, Xianglin Liu, Sirui Bi, Junqi Yin, Guannan Zhang, and Markus
  Eisenbach.
\newblock Robust data-driven approach for predicting the configurational energy
  of high entropy alloys.
\newblock {\em Materials \& Design}, 185:108247, 2020.

\bibitem{hong2022integrating}
Qi-Jun Hong, Axel van~de Walle, Sergey~V Ushakov, and Alexandra Navrotsky.
\newblock Integrating computational and experimental thermodynamics of
  refractory materials at high temperature.
\newblock {\em Calphad}, 79:102500, 2022.

\bibitem{hong2022melting}
Qi-Jun Hong, Sergey~V Ushakov, Axel van~de Walle, and Alexandra Navrotsky.
\newblock Melting temperature prediction using a graph neural network model:
  From ancient minerals to new materials.
\newblock {\em Proceedings of the National Academy of Sciences},
  119(36):e2209630119, 2022.

\bibitem{ong_python_2013}
Shyue~Ping Ong, William~Davidson Richards, Anubhav Jain, Geoffroy Hautier,
  Michael Kocher, Shreyas Cholia, Dan Gunter, Vincent~L. Chevrier, Kristin~A.
  Persson, and Gerbrand Ceder.
\newblock Python materials genomics (pymatgen): A robust, open-source python
  library for materials analysis.
\newblock {\em Computational Materials Science}, 68:314--319, 2013.

\bibitem{glushko1988thermodynamic}
Valentin~Petrovich Glushko and LV~Gurvich.
\newblock Thermodynamic properties of individual substances: Volume 1, parts 1
  and 2.
\newblock 1988.

\bibitem{russian_thermal_property_website}
Database of thermodynamic properties of individual substances.
\newblock http://www.chem.msu.su/cgi-bin/tkv.pl?show=welcome.html.
\newblock Accessed: 2023-11-09.

\bibitem{Hong2012}
Qi~Jun Hong and Axel van~de Walle.
\newblock Direct first-principles chemical potential calculations of liquids.
\newblock {\em Journal of Chemical Physics}, 137:094114, 2012.

\bibitem{hong_user_2016}
Qi-Jun Hong and Axel van~de Walle.
\newblock A user guide for {SLUSCHI}: Solid and liquid in ultra small
  coexistence with hovering interfaces.
\newblock {\em Calphad}, 52:88--97, 2016.

\bibitem{hamidieh_data-driven_2018}
Kam Hamidieh.
\newblock A data-driven statistical model for predicting the critical
  temperature of a superconductor.
\newblock {\em Computational Materials Science}, 154:346--354, 2018.

\bibitem{Supercon_dataset}
Materials data repository supercon datasheet.
\newblock https://mdr.nims.go.jp/collections/5712mb227.
\newblock Accessed: 2023-11-15.

\bibitem{zeng_atom_2019}
Shuming Zeng, Yinchang Zhao, Geng Li, Ruirui Wang, Xinming Wang, and Jun Ni.
\newblock Atom table convolutional neural networks for an accurate prediction
  of compounds properties.
\newblock {\em npj Computational Materials}, 5(1):1--7, 2019.

\bibitem{konno_deep_2021}
Tomohiko Konno, Hodaka Kurokawa, Fuyuki Nabeshima, Yuki Sakishita, Ryo Ogawa,
  Iwao Hosako, and Atsutaka Maeda.
\newblock Deep learning model for finding new superconductors.
\newblock {\em Physical Review B}, 103(1):014509, 2021.

\bibitem{gilmer2017neural}
Justin Gilmer, Samuel~S Schoenholz, Patrick~F Riley, Oriol Vinyals, and
  George~E Dahl.
\newblock Neural message passing for quantum chemistry.
\newblock In {\em International conference on machine learning}, pages
  1263--1272. PMLR, 2017.

\bibitem{he_deep_2016}
Kaiming He, Xiangyu Zhang, Shaoqing Ren, and Jian Sun.
\newblock Deep residual learning for image recognition.
\newblock In {\em 2016 {IEEE} Conference on Computer Vision and Pattern
  Recognition ({CVPR})}, pages 770--778, 2016.
\newblock {ISSN}: 1063-6919.

\bibitem{mansouri_tehrani_machine_2018}
Aria Mansouri~Tehrani, Anton~O. Oliynyk, Marcus Parry, Zeshan Rizvi, Samantha
  Couper, Feng Lin, Lowell Miyagi, Taylor~D. Sparks, and Jakoah Brgoch.
\newblock Machine learning directed search for ultraincompressible, superhard
  materials.
\newblock {\em Journal of the American Chemical Society}, 140(31):9844--9853,
  2018.

\bibitem{kaner_designing_2005}
Richard~B. Kaner, John~J. Gilman, and Sarah~H. Tolbert.
\newblock Designing superhard materials.
\newblock {\em Science}, 308(5726):1268--1269, 2005.

\bibitem{bardeen_theory_1957}
J.~Bardeen, L.~N. Cooper, and J.~R. Schrieffer.
\newblock Theory of {Superconductivity}.
\newblock {\em Physical Review}, 108(5):1175--1204, December 1957.

\bibitem{mann_high-temperature_2011}
Adam Mann.
\newblock High-temperature superconductivity at 25: {Still} in suspense.
\newblock {\em Nature}, 475(7356):280--282, July 2011.

\bibitem{bednorz1986possible}
J~George Bednorz and K~Alex M{\"u}ller.
\newblock Possible high t c superconductivity in the ba- la- cu- o system.
\newblock {\em Zeitschrift f{\"u}r Physik B Condensed Matter}, 64(2):189--193,
  1986.

\bibitem{crawshaw2020multi}
Michael Crawshaw.
\newblock Multi-task learning with deep neural networks: A survey.
\newblock {\em arXiv preprint arXiv:2009.09796}, 2020.

\bibitem{HONG2022_CMS}
Qi-Jun Hong.
\newblock Melting temperature prediction via first principles and deep
  learning.
\newblock {\em Computational Materials Science}, 214:111684, 2022.

\bibitem{HONG2022_CALPHAD}
Qi-Jun Hong, Axel {van de Walle}, Sergey~V. Ushakov, and Alexandra Navrotsky.
\newblock Integrating computational and experimental thermodynamics of
  refractory materials at high temperature.
\newblock {\em Calphad}, 79:102500, 2022.

\bibitem{MAPP}
Qi-Jun Hong.
\newblock Mapp.
\newblock
  https://faculty.engineering.asu.edu/h-ong/materials-properties-prediction-mapp/,
  2023.
\newblock Accessed: 2023-11-09.

\bibitem{MAPP_api}
Mapp-api.
\newblock https://github.com/qjhong/mapp\_api.
\newblock Accessed: 2023-11-09.

\end{thebibliography}
\end{document}